\documentstyle[aps,epsfig,12pt]{revtex}
\def\be{\begin{equation}}
\def\ee{\end{equation}}
\newcommand{\tdm}[1]{\mbox{\boldmath $#1$}}
\newcommand{\tds}[1]{\mbox{\tiny\boldmath $#1$}}
\newcommand{\unit}[1]{\mbox{\rm #1}}

\begin{document}
\draft
%\preprint{TP-JU/13-99}
\title{Atomic collisions and sonoluminescence}
\author{ Leszek Motyka$^{1}$ and Mariusz Sadzikowski$^{2}$ }
\address{ 1) Institute of Physics, Jagiellonian University,
             Reymonta 4, 30-059 Krak\'ow, Poland\\
          2) Institute of Nuclear Physics, Radzikowskiego 152,
             31-342 Krak\'ow, Poland}

\date{November, 1999}
\maketitle
\tighten

\begin{abstract}
We consider inelastic collisions between atoms of different
kinds as a potential source of photons in the sonoluminescence phenomena.
We estimate the total energy emitted in one flash and the shape
of the spectrum and find a rough agreement between the results
of our calculation and the experimental data.
We conclude that the atomic collisions might
be a candidate for the light-emitting mechanism for sonoluminescence
and discuss the implications.
\end{abstract}
\vspace*{0.3cm}

The sonoluminescence is a mechanism of the conversion of sound energy
into the energy of light emitted in a picosecond flash. This
phenomenon was discovered a long time ago (the emission of light from
the water was observed in \cite{mainesco}) but only recently has
attracted much attention because it becomes possible to trap a
single bubble of a gas mixture in a sound field \cite{gaitan} (this is
so called Single Bubble Sonoluminescence --- SBSL or just SL). The
detailed review of the topic can be found in \cite{barber}. The
mechanism of sonoluminescence has not been definitely understood so
far. There are several possible explanations, each of them possessing
its own difficulties. Let us briefly review some of the models.

One of the most promising mechanism is the Bremsstrahlung radiation from
ionised regions \cite{wu,Nat}. These regions are created by shock waves
formed during the collapse of the bubble. This model succeeds, in
particular, in prediction of the total radiation of energy per flash.
However, the presence of the plasma would imply the sensitivity of this
light-emitting mechanism to the external magnetic field \cite{young},
which is not observed.
Another possible explanation is given by the model of blackbody radiation
\cite{nolting}. The spectrum of the SL is well fitted by the blackbody
radiation spectrum of temperature 25000 K (in water of temperature
22$^o$C \cite{hiller}). However, the time dependence of SL spectrum is
independent of the color of the emitted light, which contradicts the
blackbody radiation model expectations \cite{gompag}. Other models
refer to purely quantum effects. These are: the ``dynamical Casimir
effect'' suggested by Schwinger \cite{schwinger} and the Unruh effect
elaborated in more details by Eberlein \cite{eberline}. Unfortunately,
the dynamical Casimir effect gives ambiguous predictions depending on
the renormalization procedure that one chooses, while the Unruh effect
gives the power of radiation too low by some orders of magnitude.

Yet another model describes the SL radiation as a radiation emitted
from molecular collisions \cite{frommhold}. This model properly
predicts the shape of the spectrum, however it tends to overestimate
the magnitude of the emitted energy if one uses the realistic
values of the radiating volume and the emission time.
Besides, recent developments suggest that due to
sonochemical reactions mainly noble gases are left inside the bubble
\cite{hilgenfeldt}, which if true, may be troublesome for the mechanism
of SL. Since none of the mechanisms is fully satisfactory we may,
in search for an alternative candidate, address the question
{\it if the inelastic collisions between atoms may be the source of
light in sonoluminescence}.

Thus we consider in this article the exclusive reaction
$A_1+A_2\rightarrow A_1+A_2+\gamma$ of photon production in which
the atoms remain in their ground state.
There are interesting features concerning these reactions which
encourage to study them in more detail:\\
(i) The featureless spectrum of SL suggests that the
emission of light from excited atoms do not play a crucial role
in the phenomenon.\\
(ii) The spectrum of photons emitted in atomic collisions can behave
similarly to that of SL. Indeed, the emission of photons
with a wavelength greater than the size of the atom is suppressed
due to the mutual cancellations between the contributions of
the nucleus and the electronic cloud (c.f. the Rayleigh scattering
process). The short wavelength photons are absent because of
the cut-off on the energy available in the collision.

In the case of SL from an air bubble in water the analysis of
the scattering of the oxygen or nitrogen atoms on the atoms
of argon should be performed. The exact calculation for this process
is extremely involved thus, using a simplified model,
we estimate the order of magnitude of the cross-section and the shape
of the spectrum. Hence, let us start from the process of scattering of
a hydrogen atom in the Coulomb field of an infinitely heavy source,
accompanied by emission of light.
After obtaining the cross-section we shall argue that this elementary
process has similar features to the atomic collision and
we shall use the results to estimate the number of photons
in the SL flash.

We employ the standard non-relativistic hamiltonian whose leading
part $H_0$ contains the operators of the kinetic energy and the
electron-proton interaction. The perturbation $H_1$ in the Coulomb
gauge, after neglecting $\tdm A^2$ terms, take the following form:
\be
H_1 (\tdm r_e, \tdm r_p) = - {e\tdm A (\tdm r_e)
\tdm p_e \over m_e} + { e\tdm A (\tdm r_p) \tdm p_p \over m_p}
- {Ze^2 \over r_e} + { Z e^2 \over r_p},
\ee
where $\tdm r_e$~($\tdm r_p$) and $\tdm p_e$~($\tdm p_p$) are the
position and momentum of the electron (proton), the source of the potential
is situated in the origin and carries the electric charge of $Ze$,
$\tdm A_k (\tdm r) =
\tdm \varepsilon_k {e^{i\tds k \tds r} \over \sqrt{V}}
a^+ _{\varepsilon, k}$
is the vector potential of the emitted plain wave with momentum $\tdm k$ and
the polarisation vector given by $\tdm \varepsilon_k$.
The matrix element of the creation operator in the box is given by
$\langle \gamma(\varepsilon,\tdm k)| a^+ _{\varepsilon,k} |0\rangle
= \sqrt {2\pi \over k}$.
The eigenstates of $H_0$ are given in the position space by
the wave functions
$
\Psi_{P, n} (\tdm R, \tdm r) =
{1 / \sqrt{V}} \exp(-i\tdm P \tdm R) \,\psi_n (\tdm r)
$
with $\tdm P$ being the atom momentum, $\psi_n (\tdm r)$ ---
the electron-proton wave function (for a bound or continuum state $n$)
in the centre-of-mass frame and
$\tdm r = \tdm r_e - \tdm r_p$,
$\tdm R = \alpha \tdm r_p+ \beta \tdm r_e$,
$\alpha = m_p /(m_e + m_p)$,
$\beta = m_e /(m_e +m_p)$.
For future use, we also introduce
the mass of the atom $M_H=m_e+m_p$. We shall calculate the amplitude 
${\cal M}$
of the transition: $\Psi_{P, 1S} \to \Psi_{P', 1S} \gamma (\tdm k)$ in
the second order of perturbative expansion which gives the first
non-vanishing contribution. In this approximation the amplitude reads
\be
%{\cal M} (\Psi_{P,1S} \to \Psi_{P',1S} \gamma (\tdm k) ) =
{\cal M} =
\sum _{m} {\langle \Psi_{P',1S} \gamma (\tdm k) | H_1 | m\rangle
           \langle m | H_1 | \Psi_{P, 1S} \rangle \over
                E_{P,1S} - E_m + i\epsilon},
\label{master}
\ee
where $m$ runs over all the possible intermediate states.
There are two groups of the virtual states depending on the photon
content. The first one contains
only the electron-proton states moving with the total momentum
$\tdm P''$, which may be either virtual bound states or virtual
continuum states. We shall denote the members of this group by
$|\Psi_{P'',n}\rangle$. The other group includes the states
$|\Psi_{P'',n}; \gamma(\tdm k)\rangle$ built of a photon of energy
$k$ and momentum $\tdm k$ accompanying states $|\Psi_{P'',n}\rangle$
in the electron-proton sector.
It is straightforward to observe,
that the sum (\ref{master}) may be expanded into two parts corresponding
to final (FSR) and initial state (ISR) photon radiation amplitudes.
%It is straightforward to observe,
%that the sum (\ref{master}) may be expanded as follows
%\[
%{\cal M} (\Psi_{P,1S} \to \Psi_{P',1S} \gamma (\tdm k) ) =
%\sum _{n,P''} \left[ {
%        \langle \Psi_{P',1S} \gamma (\tdm k) | H_E | \Psi_{P'',n} \rangle
%        \langle \Psi_{P'',m} | H_S | \Psi_{P,1S} \rangle
%        \over
%        E_{P,1S} - E_{P'',n} + i\epsilon} +
%        \right.
%\]
%\be
%        \left.
%        {\langle \Psi_{P',1S} \gamma (\tdm k) |H_S|\Psi_{P'',n}
%                                       \gamma (\tdm k) \rangle
%        \langle \Psi_{P'',m} \gamma (\tdm k)| H_E | \Psi_{P,1S}\rangle
%        \over
%        E_{P,1S} - E_{P'',n} - k + i\epsilon}
%         \right]
%\label{master2}
%\ee
%where $H_S$ and $H_E$ are the components of $H_1$ responsible for
%the atom Coulomb scattering and for the photon emission respectively.
%These two parts correspond to final (FSR) and initial state (ISR) photon
%radiation amplitudes.
>From the kinematics of the process and the conservation
of momentum it is easy to find that the energy denominators
take the following form:
$
D_{\small\rm FSR} \simeq E_{1S} - E_n + k
$
and
$
D_{\small\rm ISR} \simeq E_{1S} - E_n - k
$
where, making the approximations, we have employed the hierarchy
$k \ll P \sim P' \ll M_H$.

%The sum (\ref{master2}) is still too complicated to be evaluated
%analytically thus we shall make further approximation.
%
We have checked, that the contributions to the amplitude
%${\cal M} (\Psi_{P,1S} \to \Psi_{P',1S} \gamma (\tdm k) )$
${\cal M}$ of the intermediate states in which the electron and proton
form a $1S$ state are negligible, due to $a_0 ^2 k^2$ suppression
of the photon emission matrix elements.
%
%$\langle \Psi_{P'',1S} \gamma (\tdm k)| H_E | \Psi_{P, 1S} \rangle$ and
%$\langle \Psi_{P',1S}\gamma (\tdm k) | H_E | \Psi_{P'', 1S} \rangle$.
%
On the other hand, the spectrum of hydrogen atom has a large gap
between the ground state and the first excited state and the
excited bound states lie close to each other. It also may be proven
that the intermediate continuum states with energies
$E_n > |E_{1S}|$ give a subleading contribution to the amplitude.
It seems therefore quite natural to approximate the energy difference
$E_{1S} - E_n$ by a characteristic energy, say
$E_{1S} - E_n \simeq -\Delta E = -|E_{1S}|$.
Of course, this approximation works only for $k$ significantly
smaller than the $\Delta E$.
Now, we can employ the completness relation for the $\Psi_{P,n}$ states
in the electron-proton sector and perform the sum over $P,n$
to obtain
\footnote{
This approach is in the spirit of the classical approximation
applied in the calculation of van der Waals potential \cite{VdW}. }
\be
%{\cal M} (\Psi_{P,1S} \to \Psi_{P',1S} \gamma (\tdm k) ) \simeq
{\cal M} \simeq
{\langle \Psi_{P',1S} \gamma (\tdm k) | H_S H_E | \Psi_{P, 1S} \rangle \over
                -\Delta E-k} +
{\langle \Psi_{P',1S} \gamma (\tdm k) | H_E H_S | \Psi_{P, 1S} \rangle \over
                -\Delta E+k},
\ee
where $H_S$ and $H_E$ are the components of $H_1$ responsible for
the atom Coulomb scattering and for the photon emission respectively.
After performing the necessary integrations and taking into account
the relation $\tdm \varepsilon_{k}\, \tdm k = 0$ we get
the following expression for the amplitude:
\[
%{\cal M} (\Psi_{P,1S} \to \Psi_{P',1S} \gamma (\tdm k) ) \simeq
{\cal M} \simeq
- \sqrt{2\pi \over k V^3} {4 \pi Z e^3 \over
(\tdm P' + \tdm k - \tdm P)^2  } \times
\]
\[
\left\{
{k \over \Delta E ^2 - k^2}\;\tdm\varepsilon_k \tdm q \;
\left[
{\beta \over m_e} f(\beta\tdm q + \tdm k)
+ {\beta \over m_p} f(\beta\tdm q)
+ {\alpha \over m_e}  f(\alpha\tdm q)
+ {\alpha \over m_p}  f(\alpha\tdm q-\tdm k)
\right] +
\right.
\]
\be
\left.
{2\Delta E \over \Delta E^2 - k^2}\; \tdm\varepsilon_k (\tdm P + \tdm P') \;
{1 \over M_H}  \;
\left[ f(\alpha \tdm q) - f(\alpha \tdm q + \tdm k) +
       f(\beta \tdm q) - f(\beta \tdm q + \tdm k)
\right]
\right\}
\ee
where $\tdm q = \tdm P' - \tdm P$ and
$f( \tdm l) = (1 +  a_0 ^2 l^2 /4)^{-2} $
is the electric form-factor of the ground state of the hydrogen atom
of the Bohr radius given by $a_0$.
As expected, there occur cancellations between the form-factors,
and the amplitude vanishes for vanishing $k$.
We also notice that the terms proportional to
$ \tdm\varepsilon_k (\tdm P + \tdm P')$ are suppressed by
a small factor $a_0 k$ as compared to term containing
$ \tdm\varepsilon_k \tdm q $ thus we retain only the later.
Furthermore we may drop the $1/m_p$ terms.

A more subtle problem arises when comparing the relative importance
of the contributions
${\beta \over m_e} f(\beta\tdm q + \tdm k)$ and
${\alpha \over m_e} f(\alpha\tdm q )$.
The answer seems to depend on the details of the process in the
physically relevant parameter space.
For example, if the energy of the collision falls between 5 and~10~eV
then the first term dominates for $k>3$~eV while for $k<3$~eV
the contribution of the other term is more important giving rise
to an additional ``red'' peak of spectral density. However, if
we take an atom with the mass of $16~M_H$ and the Bohr radius $a_0$
(modelling the oxygen atom) this peak would move to energies
$k<0.5$~eV i.e. to the experimentally inaccesible region.
Besides, the ${\alpha \over m_e} f(\alpha\tdm q )$ term is very sensitive
to the details of the charge form-factor of the scattering atom.
On the other hand, the ${\beta \over m_e} f(\beta\tdm q + \tdm k)$ part
is universal i.e very weakly dependent on the details of the atom structure.
In what follows, we shall focus only on this universal contribution
of the amplitude which is relevant for the shape of the spectrum
in its large $k$ part. By this neglection we underestimate the number of
the emitted photons.

Thus we get the estimate
\be
%{\cal M} (\Psi_{P,1S} \to \Psi_{P',1S} \gamma (\tdm k) ) =
{\cal M} =
-\sqrt{2\pi \over k V^3} {4 \pi Z e^3 \over M_H }
{k \over \Delta E^2 - k^2}\; {\tdm\varepsilon_k \tdm q \over q^2}
\label{amplitude}
\ee
>From this amplitude, after performing the standard integrations over the
phase space, one gets the differential cross-section for the photon
emission from a collision at the energy $E$:
\be
{d\sigma\over dk} =
{8 Z^2 e^6  \over 3 M_H E} {k^3 \over (\Delta E^2-k^2)^2}\;
\log \left( {P+\bar P \over P - \bar P} \right),
\label{dsdk}
\ee
where
$P = \sqrt{2M_H E}$, $\bar P = \sqrt{ 2M_H (E-k)}$.
For the purpose of the numerical calculations, we take
$\Delta E =13.6$~eV.

Let us point out, that the obtained cross-section corresponds to a
subprocess in which the nucleus scatters off the Coulomb field and the
electron cloud radiates a photon when recombining around the scattered
nucleus. The other contributions to the amplitude were shown to be
subleading and neglected. The spectral density resulting from this
cross-section rises as $k^4$ for small $k$ i.e faster than this
observed in the SL experiments. For $k$ closer to the
kinematical cut-off, the initial $k^4$ dependece is modiffied and one
gets approximately quadratic behavior over wide range of $k$ which is
still steeper that it follows from the experimental data. However, let
us remind, that in our picture, the SL spectrum would be given by a
convolution of the spectra of individual collisions with the (unknown)
distribution of the collision energy, so the two spectra may differ
from each other.

After integration of the differential cross-section (\ref{dsdk}) over
the energy, we get the total cross-section $\sigma_{tot}(E)$ growing
with the collision energy $E$ approximately as $E^\nu$ with
$\nu \simeq 3.5$. For the choice of parameters relevant for the
hydrogen atom, $Z=1$ and a typical collision energy of $7~\unit{eV}$
(the choice is suggested by the observed cut-off on photon spectrum)
we obtain $\sigma_0 = \sigma_{tot}(7~\unit{eV})=1.3\cdot
10^{-31}$~m$^2$. Taking this number and the parameters implied by the
known facts concerning the bubble dynamics we may, after some
modifications, make a crude estimate of the number $N_f$ of photons
produced in one flash. We assume, that the bubble is filled with a
mixture of atoms of a noble gas and atoms of other element (oxygen or
nitrogen)
\footnote{At the temperatures and densities predicted by a
shock-wave model the diatomic molecules are dissociated.}
with atomic numbers $Z_N$ and $Z_O$, masses $M_N$ and $M_O$ and
concentrations $n_N$ and $n_O$ correspondingly.
The predictions for collisions of an atom of noble gas with an
atom of, say, oxygen may be formulated on the basis of the previous
calculation. Namely, for the momentum transfers relevant for SL
the condition $a_0^2 q^2 \gg 1$ is fulfilled and
the charge form-factors for the electrons
suppress their contribution to the scattering amplitude  --- the
electrons cannot absorb momenta much larger than
their average momentum in the atom.
Then the nucleus of the noble atom acts as a bare source of the Coulomb
field.
Furthermore, the amplitude of the electromagnetic radiation from the noble atom
is smaller then from the oxygen.
It is caused by the smaller polarizability of the
former atom which is governed by the relevant ionisation energy.
Thus the cancellations which occur
due to the destructive interferece do not reduce the cross-section
substantially in contrast with the case
of a collision of two objects of the same kind.
This feature was reflected in our model by neglecting the
radiation from the source of the Coulomb field.
Of course, to complete the analogy, we should use in the formulae the
reduced mass of the two atoms instead the mass of the hydrogen
atom and include the modification of charges by $Z_N$ for the source
of the field and $Z_O $ for the scattering atom. The charge
$Z_O$ enters in fourth power since it contribute both to scattering
and the radiation. Therefore we approximate $N_f$ by the following
expression:
\be
N_f \simeq {4\pi R_s^3\over 3}\,v\tau \;
{M_H(M_O+M_N) \over M_O M_N } Z_N^2 Z_O ^4 \, \sigma_0\;  n_N n_O,
\label{Nf}
\ee
where $R_s$ is the radius of the hot gas region,
$\tau = 100$~ps is the light emission time
\cite{gompag} and $v \simeq 10^4$~m/s denotes the relative velocity
of colliding atoms. Let us focus on the argon ($Z_N=18$)--oxygen ($Z_O=8$)
process, which may be relevant for the SL of air bubble in water.
As a first guess we take for $R_s$ the
minimal radius of the bubble i.e. about 0.5~$\mu$m and assume
for the concentrations $n_O + n_N \simeq  600 n_0$ in accordance
with the measured compression factor \cite{barber}, with
$n_0 = 2.7\cdot 10^{25}$~m$^{-3}$ being the concentration of the ideal
gas in the normal conditions. We obtain $N_f \simeq 5\cdot 10^5$
for $n_O = n_N$, in a reasonable agreement with the experiment.
In the alternative shock-wave description of gas dynamics lower values of the
radius are prefered, however the increase in concentration caused by
the additional compression may easily compensate the decrease of the
reaction volume. In fact, the true concentration in the radiating region
is not measured, and the models provide us only with ambigous predictions.
The limitting value of the concentration is
$\sim 1/(8 a_0^3)=10^{30}$ m$^{-3}$
thus there is still some room here and the constraints imposed on the
cross-section (\ref{Nf}) by the experiment are not very stringent.
It is also very probable, that the proposed mechanism is not responsible
for the emission of all photons but rather supplements the list
of processes studied in \cite{Nat}. In this case, including it
could, in particular, improve the shape of the spectrum obtained in
\cite{Nat}.

We can also take into account that, as follows from \cite{hilgenfeldt},
the SL bubble is filled mainly with a noble gas, and may contain only
some admixture of oxygen (which may be continously provided from water)
by modifying the ratio $n_O:n_N$, keeping $n_N+n_O$ fixed. Thus
for 10\% of the oxygen in the gas mixture the
number of photons in one flash drops to about $2\cdot 10^5$.

In conclusion, we propose a novel light-emitting mechanism in
sonoluminescence, in which the photons are radiated from
atoms disturbed by collisions. The number of produced photons
has the proper order of magnitude, however the resulting photon
spectrum comes out somewhat to steep. The spectral density
is featureless and universal. Our model of SL requires the gas
temperature to be approximately 30~000~K, and the concentration
of the order of $500-1000~n_0$. The bubble should contain atoms of two
different gases: a noble gas and a gas with smaller ionisation energy
e.g. oxygen, however one of the gases may appear at much smaller
concentration than the other. The emitting process is insensitive
to external magnetic field and does not require the presence of molecules
which are expected to dissociate in the gas temperature suggested by
the photon spectrum.
Although the model we have elaborated is based on the simplified
assumptions we expect that the obtained predictions estimate
correctly the order of magnitude of the cross-section.
Therefore this gives the strong
motivation to further study of the subject which
would probably require the use of numerical analysis if one intends to
cover all the details of the process.

\section*{Acknowledgements}
We are very grateful to Professors W.~Czy\.{z} and K.~Zalewski for useful
comments and to L.~Hadasz and B.~Ziaja for helpful discussions.
This research was supported in part by the Polish State Committee
for Scientific Research (KBN) grants 2P~03B~084~14 and  2P~03B~086~14.

\end{document}